# Conséquences du changement climatique pour les maladies à transmission vectorielle et impact en assurance de personnes

**Yannick Drif[1], Benjamin Roche[2] et Pierre Valade[1, 3]**

[1] Département Vie, Accident et Santé, Aon France, Paris, France
[2] UMR IRD 224-CNRS 5290, Université de Montpellier, Montpellier, France
[3] Actuaire Certifié, Expert ERM de l'Institut des Actuaires Français

E-mail : yannick.drif@aon.com ; pierre.valade@aon.com

**Résumé**

Le changement climatique, grandement relié aux activités humaines, a d'ores et déjà un impact considérable sur nos sociétés. Au vu des tendances actuelles, le changement climatique devrait s'accélérer durant les prochaines décennies. Au-delà de son impact sur le rythme des catastrophes naturelles (inondations, ouragans, etc…), le changement climatique peut avoir des conséquences catastrophiques pour la vie et la santé humaine. Une des préoccupations avérées est l'augmentation de la transmission des virus diffusés par les moustiques. En effet, l'augmentation des températures a un impact positif direct sur la viabilité du moustique dans les écosystèmes, ce qui conduit à son abondance et donc au risque d'exposition des populations humaines à ces pathogènes.

Cette étude quantifie les conséquences du réchauffement climatique sur le risque d'épidémies de virus transmis par le moustique Aedes Albopictus en France Métropolitaine. Ce moustique, vecteur, entre autres, des virus de la Dengue, du Chikungunya ou du Zika, est arrivé sur le territoire métropolitain en 2004 et s'étend depuis à tout l'Hexagone. Grâce à l'association précédemment établie entre la probabilité de la présence du moustique et la température moyenne combiné avec un modèle mathématique, la probabilité d'épidémie, le nombre de personnes pouvant être infectées et pouvant décéder durant une saison dans chaque département sont estimées. S'il existe une hétérogénéité forte sur le territoire métropolitain, près de 2,000 décès par an pourrait être à déplorer à l'horizon 2040.

Mots clés : dengue, moustique, pandémie, vecteur, climat, assurance

## 1. Changement Climatique et transmission vectorielle

En 1992, l'Union of Concerned Scientists (1 700 signataires comprenant les plus grands scientifiques du monde et de nombreux des lauréats de prix Nobel) a publié une tribune intitulé World Scientists Warning to Humanity (l'avertissement des scientifiques du monde à l'humanité) dans lequel ils soutenaient que les activités humaines étaient susceptibles de conduire à une catastrophe écologique, sanitaire et sociale (Union of Concerned Scientists 1993).

Depuis cet avertissement, les tendances environnementales observées n'ont fait que s'aggraver. Un second volet de cette tribune, publié en 2017 (Ripple et al. 2017), fait un état des lieux de ces tendances observées entre 1992 et 2016, pour 9 grands indicateurs environnementaux. Ces résultats suggèrent une réduction significative de la biodiversité, une augmentation de la déforestation, de l'émissions mondiales de carbone et des températures moyennes… qui menacent considérablement le bien-être humain.

### *1.1 Modèles climatiques*

La majeure partie des débats politiques actuels concernent plus spécifiquement les mesures à prendre pour limiter le changement climatique. Selon le groupe d'experts intergouvernemental sur l'évolution du climat (GIEC ou IPCC en anglais (The Intergovernmental Panel on Climate Change n.d.)), le changement climatique correspond à une modification durable du climat à l'échelle planétaire due à une augmentation des concentrations des gaz à effet de serre (CO2, mais aussi méthane par exemple) dans l'atmosphère. Il est principalement lié à des causes naturelles comme l'effet des cycles solaires, les éruptions volcaniques ou même la dérive des continentsfortement relié à l'activité humaine depuis la révolution industrielle (moitié du XIXe siècle,(IPCC 2014)).

Il est donc primordial d'estimer l'ampleur du changement climatique (actuel et futur) et quelles seront ses conséquences à l'échelle régionale et mondiale. Pour cela, des modèles climatiques s'appuient sur les données récoltées lors des 150 dernières années (tel que les mesures de température ou les émissions de gaz à effet de serre) pour fournir les informations nécessaires à l'évaluation des impacts du changement climatique sur la santé humaine ou encore la perte de biodiversité. Grace à ces modèles, les scientifiques établissent des scénarios futurs pour élargir notre champ de vision à l'horizon 2100. Ces études démontrent que, même si nous utilisons environ 15% des ressources existantes, des modifications conséquentes de notre environnement seront à prévoir.

La dernière génération en date de modèle climatique (programme mondial de simulations du climat CMIP6 (CNRS n.d.)) sera utilisée dans le prochain rapport du GIEC qui devrait être publié en 2021. Deux des modèles impliqués dans le programme CMIP6 ont été développés par des groupes français, l'un par le CNRS et Météo-France (CNRM), l'autre par l'Institut Pierre-Simon-Laplace (IPSL) (Figure 1). Ces deux modèles s'accordent à dire que le scénario le plus pessimiste (une croissance économique rapide alimentée par des énergies fossiles - SSP5 8.5) mène vers une augmentation de la température moyenne globale de l'ordre de 6 à 7°C en 2100, soit une augmentation plus élevée que les estimations du précédent rapport du GIEC (CMIP5 suggéré que la température mondiale moyenne pourrait augmenter de 2,0°C d'ici 2100(Taylor et al. 2009)). Le scénario le plus atténué (supposant l'application de politiques de réduction drastique, une forte coopération internationale qui donnerait priorité au développement durable - SSP1 1.9), permet de rester sous l'objectif des 2°C de réchauffement, au prix de gros efforts sur les activités humaines. Ces études fournissent donc des informations cruciales sur l'évolution à long terme du climat planétaire et alimentent le débat public sur les mesures à prendre.

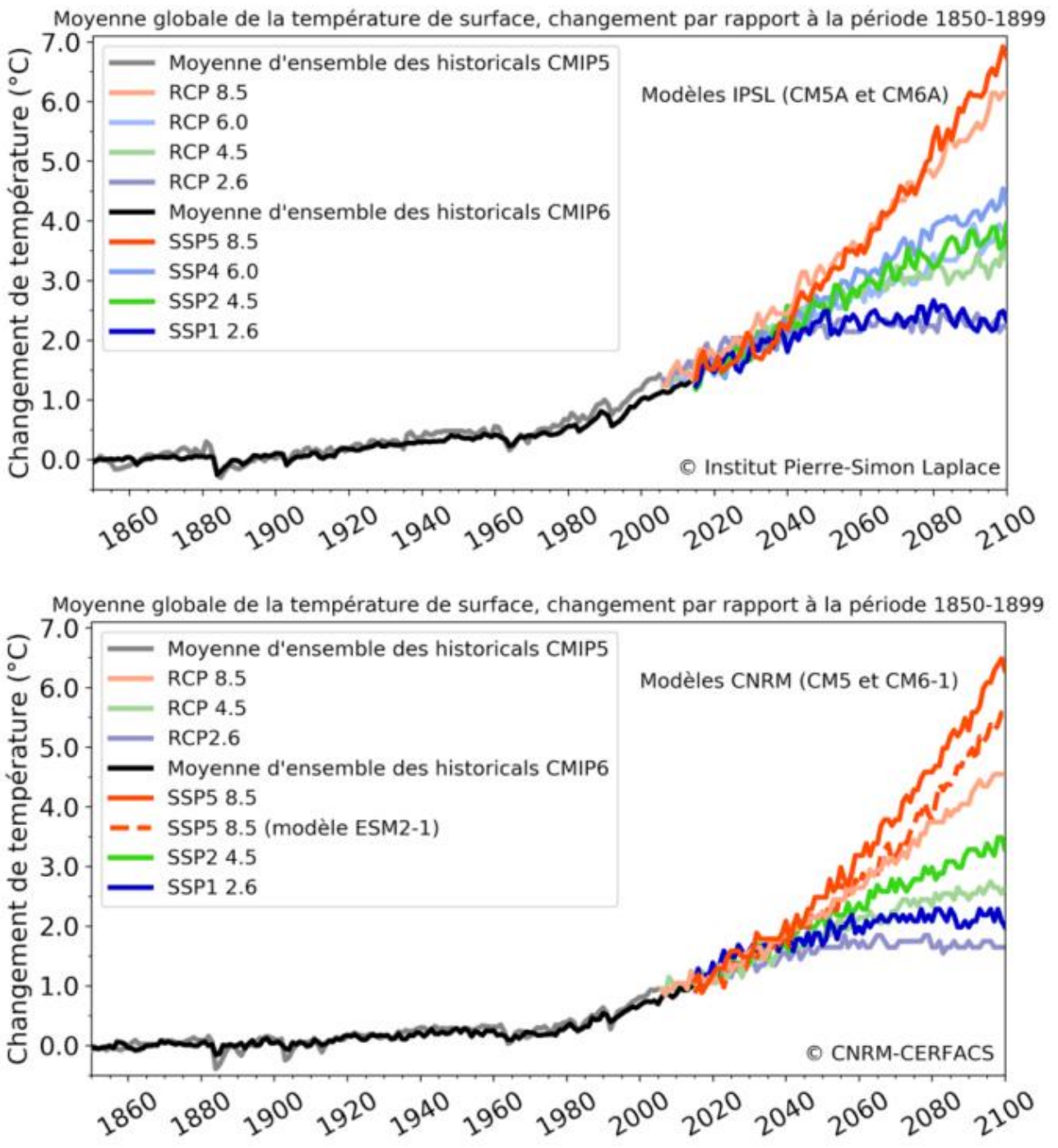

***Figure 1**. Évolution des températures moyennes de la planète simulées pour le programme CMIP5 publié en 2012 (traits de couleur pastel) et pour le programme CMIP6 (traits de couleur vive) pour les deux modèles français l'un développé par le CNRS/Météo-France (CNRM), l'autre par l'Institut Pierre-Simon-Laplace (IPSL). Source : CNRS (CNRS n.d.).*

## *1.2 Impact sur le cycle de vie des moustiques*

Un phénomène constaté de ce changement climatique sur la santé humaine est la distribution des maladies infectieuses à l'échelle mondiale (Lafferty 2009, Hoberg and Brooks 2015, Escobar et al. 2016). En particulier, les maladies transmises par des arthropodes vecteurs contribuent largement au fardeau planétaire des maladies infectieuses (WHO 2017). A chaque instant, près de la moitié de la population mondiale est exposée (i.e., risque d'être infecté) à un agent pathogène transmis par un vecteur (WHO 2014).

Les moustiques constituent un groupe clef de préoccupation en santé publique puisqu'ils transmettent d'importantes maladies émergentes tel que la Dengue (environ 96 millions de cas par an), le Chikungunya (environ 693 000 cas par an) ou encore la maladie à virus Zika (environ 500 000 cas par an). Au-delà des conséquences sanitaires de ces maladies à la morbidité et la mortalité élevées (Kilpatrick and Randolph 2012, Franklinos et al. 2019, Ryan et al. 2019), Tableau 1), elles peuvent également impacter le développement économique et sociétal d'un pays.

***Tableau 1**. Nombre de cas et de décès annuel à l'échelle mondiale des principales maladies transmises par les moustiques d'importance en santé publique. Source : (WHO 2014, 2017, Franklinos et al. 2019)*

| Maladies infectieuses | Vecteurs prédominants par genre | Nombre de cas annuel estimé ou signalé | Nombre de décès estimé |
|---|---|---|---|
| Chikungunya | Aedes, Anopheles, Culex et Mansonia | 693 000 (Continent Américain) | 2 000 par an |
| Dengue | Aedes | 96 million (± 67 - 136 million) | 25 000 par an |
| Encéphalite japonaise | Culex | 42 500 (± 35 000 - 50 000) | 13 600 à 20 400 par an |
| Fièvre jaune | Aedes et Haemagogus | 130 000 (± 84 000 - 170 000, Continent Africain) | 30 000 par an |
| Paludisme | Anopheles | 212 million (± 148 - 304 million) | 500 000 par an |
| Virus du Nil occidental | Culex | 2 588 | Quelques dizaines par an en moyenne |
| Virus Zika | Aedes | 500 000 (Continent Américain) | 96 000 depuis 2015 |

Les connaissances actuelles suggèrent que l'impact des maladies transmises par les moustiques devrait s'étendre considérablement en conséquence du changement climatique (Messina et al. 2015, Franklinos et al. 2019). En effet, la dynamique de transmission des agents pathogènes (WHO 2004) mais aussi le développement et la survie des moustiques sont fortement influencés par les facteurs climatiques. La température et les précipitations sont tout particulièrement importantes.

Les changements de température affectent, par exemple, le temps nécessaire au développement des agents pathogènes (période d'incubation) dans le moustique (Paaijmans et al. 2012, Mordecai et al. 2017). Des études ont montré que cette période d'incubation se raccourcit avec l'augmentation de la température, augmentant par la période durant laquelle ces agents pathogènes sont susceptibles de transmettre les maladies (Reisen et al. 2006, Barbazan et al. 2010, Tjaden et al. 2013). Le réchauffement global affecte également les paramètres biologiques du moustique en diminuant, par exemple, la durée de son cycle gonotrophique (le temps nécessaire au moustique à la digestion d'un repas de sang et à la maturation de ses œufs entre deux repas sanguins) (Barbazan et al. 2010). Le changement climatique mène ainsi à une accéléreration des cycles de reproduction des moustiques. Cela augmentera tant le nombre de moustiques que la probabilité de transmission des agents pathogènes par ces espèces (Jetten and Focks 1997, WHO 2004).

Il est important de préciser que, dans les zones tempérées, la saisonnalité est une composante clé à la survie et de la diffusion de certains agents pathogènes (Paul 2012). En effet, les températures estivales y sont au moins aussi élevées que dans les saisons les plus chaudes de la plupart des zones tropiques. La différence cruciale est que les tropiques ne subissent pas de saisons froides. Si des agents pathogènes tropicaux, transmis par les moustiques, sont introduits à la bonne saison dans une zone tempérée, ils peuvent être diffusés par les vecteurs appropriés qui y sont présents. Dans la plupart des cas, ils seront éliminés à l'arrivée de l'hiver. Dans ce contexte, le changement climatique devrait avoir un effet sur cette saisonnalité en racourcissant les périodes froides pendant l'hiver. Les périodes de transmission de ces pathogènes seront alors augmentée en durée.

### *1.3 Les maladies vectorielles – Un nouveau danger en France Métropolitaine*

Les virus transmis par ces arthropodes vecteurs (arbovirus) représentent une nouvelle menace en France. En 2015, près de 1,000 cas importés de Dengue et Chikungunya ont été détectés en Métropole. Des personnes, ayant contracté le virus dans des pays tropicaux, reviennent sur le territoire français avec la possibilité de transmettre le virus par le biais de piqures de moustiques. Ces derniers servent alors de vecteur et contaminent d'autres personnes sur le territoire français.

Jusqu'à récemment, ce risque était négligeable car aucun moustique capable de transmettre ces virus n'était recensé en Métropole. Néanmoins, à partir de 2004, le moustique Aedes Albopictus, communément surnommé moustique tigre, à commencé à s'implanter en France Métropolitaine (Roche et al. 2015). Venant d'Italie, sa propagation semble surtout le résultat d'un transport passif de moustiques adultes au travers de déplacement en voitures ou en camion. Cela est cohérent avec le constat de sa propagation, qui suit parfaitement le réseau autoroutier. Néanmoins, sa présence est aussi dépendante d'autres facteurs abiotiques, notamment le type de paysage ou encore la température locale.

En conséquence, des épisodes de transmission autochtone (i.e., transmission entre individus sur le territoire métropolitain) ont déjà eu lieu. Ainsi, 12 cas de Chikungunya ont été détectés à Montpellier en 2014 ou 7 cas de Dengue à Nîmes (Delisle et al. 2015, Succo et al. 2016). Ces épisodes se sont répétés ces dernières années dans le sud de la France, mais ont toujours été relativement rapidement contrôlés grâce à un dispositif de lutte anti-vectorielle local efficace. En revanche, le moustique tigre s'étend vers le Nord de la France où peu de services de démoustication existent, ce problème pourrait devenir s'intensifier dans les années à venir.

## 2. Modélisation

### *2.1 Méthodologie*

Le scénario RCP8.5 (surnommé business as usual) a été considéré à 3 horizons différents : 2025, 2030 et 2040. Ce scénario de réchauffement climatique prédit une évolution de la température de 0,5° en 2025, 1° en 2030 et 2° d'ici à 2040.

Grâce à l'étude précédemment réalisées sur cette espèce de moustique (Roche et al. 2015), nous avons pu mettre en évidence qu'une augmentation de 1 degré de la température locale au mois de janvier (mois le plus froid de l'année) augmente la probabilité de présence du moustique de 7 %. Néanmoins, la température doit être supérieure à 10.4° pour que cette présence soit non nulle.

En se basant sur la proportion de la population humaine exposée au virus (Direction Générale de la Santé n.d.), chaque département été donc classés en présence haute ou basse du moustique représentant la proportion de la population exposée au moustique (considéré ici à 50 % 20 % respectivement). Ensuite, la proportion de la population exposée au moustique, qui est directement reliée à la probabilité de présence du moustique, est considérée augmenter de 3.5 %, 7 % et 14 % en 2025, 2030 et 2040 respectivement si la température est supérieure à 10.4°.

Les proportions de la population locale exposée au moustique dans chaque département en 2020, 2025, 2030 et 2040 sont ensuite inclus au sein d'une version simplifiée d'un modèle mathématique (Sochacki et al. 2016) Où les populations humaines et de moustiques sont divisées selon leur statut infectieux.

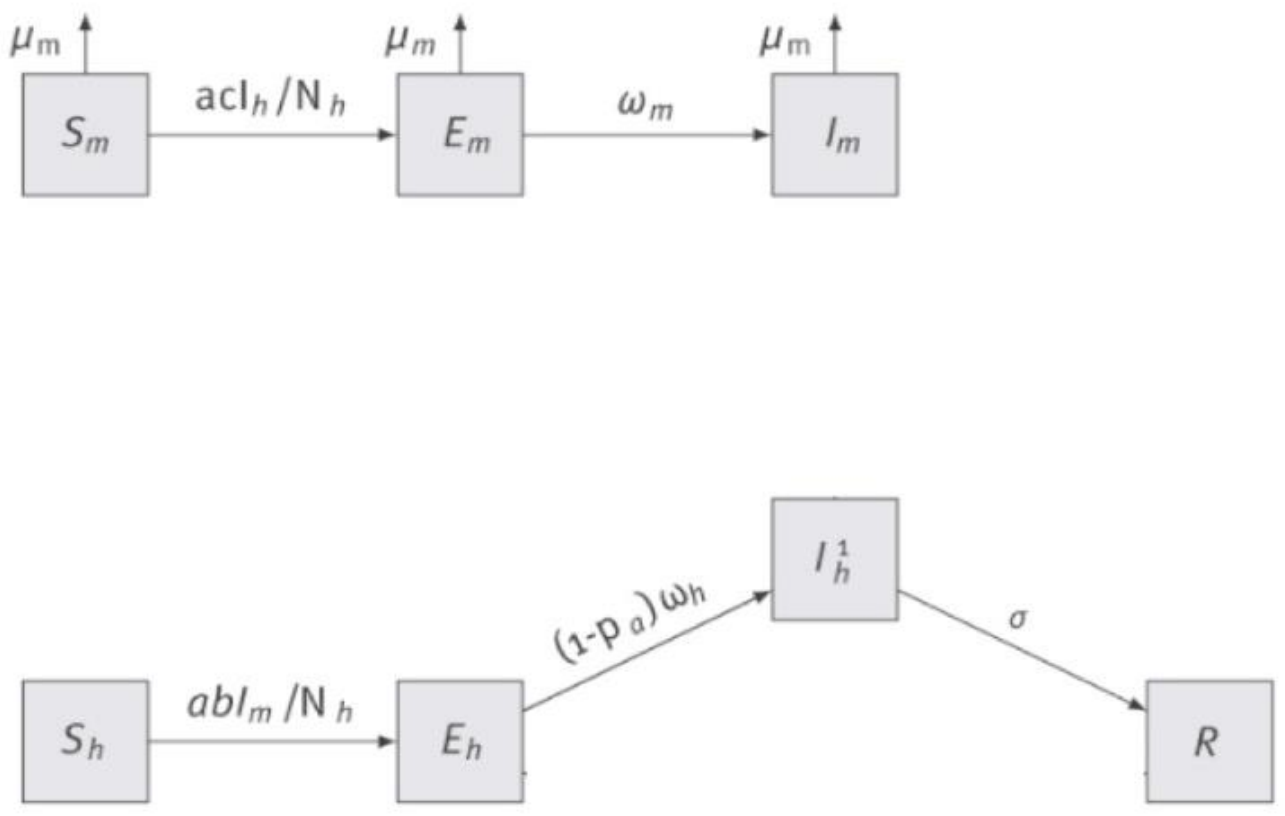

***Figure 2**. Modèle mathématique de transmission d'une arbovirose utilisé au sein de ce rapport. Source : (Sochacki et al. 2016)*

Les individus commencent dans l'état Susceptible (*Sh* et *Sm* pour l'homme et le moustique respectivement), où ils peuvent être infectés. Ils deviennent ensuite exposés (*Eh* et *Em* pour l'homme et le moustique respectivement), c'est-à-dire infectés mais pas encore infectieux. Après une période de latence, ils deviennent infectieux (*Ih* et *Im* pour l'homme et le moustique respectivement) et peuvent cette fois-ci transmettre le virus à l'homme (pour les moustiques infectieux *Im*) ou au moustique (pour les humains infectieux *Ih*). Les paramètres du modèle sont les mêmes que ceux utilisés dans l'article (Sochacki et al. 2016).

Comme la transmissibilité du virus par la population de moustique en France Métropolitaine est peu connue, deux hypothèses ont été utilisées. Nous avons considéré une hypothèse de transmission basse avec un R0=1.1, et une hypothèse de transmission forte avec un R0=2.2. Le R0 représente le nombre de cas secondaires qu'un cas primaire peut créer au sein d'une population entièrement susceptible (Keeling et al. 2008) et peut facilement être calculé avec le modèle mathématique précédemment exposé.

Enfin, les simulations sont réalisées avec une version stochastique du modèle suivant l'algorithme du *tau-leap* (Keeling et al. 2008). Chaque simulation est centrée sur un département, et le nombre d'individus infectés au début de la simulation est égal à 1 000 (i.e., le nombre d'individus infectés par un arbovirus arrivant en France chaque année) multipliée par la représentativité de la population du département par rapport à la population française (i.e., posant comme hypothèse que les départements les plus peuplés recevront le plus de cas importés). Pour chaque département et chaque scénario, 100 réplicas ont été réalisés.

## 2.2 *Résultats*

### *2.2.1 Probabilité d'apparition d'épidémies*

On s'intéresse d'abord à la probabilité de survenue d'une épidémie dans chaque département. Pour cela, sur les 100 simulations réalisées pour chaque département et chaque scénario climatique, le nombre de simulations ayant donné au moins une fois une transmission locale est calculé, donnant la probabilité de survenue d'une transmission autochtone.

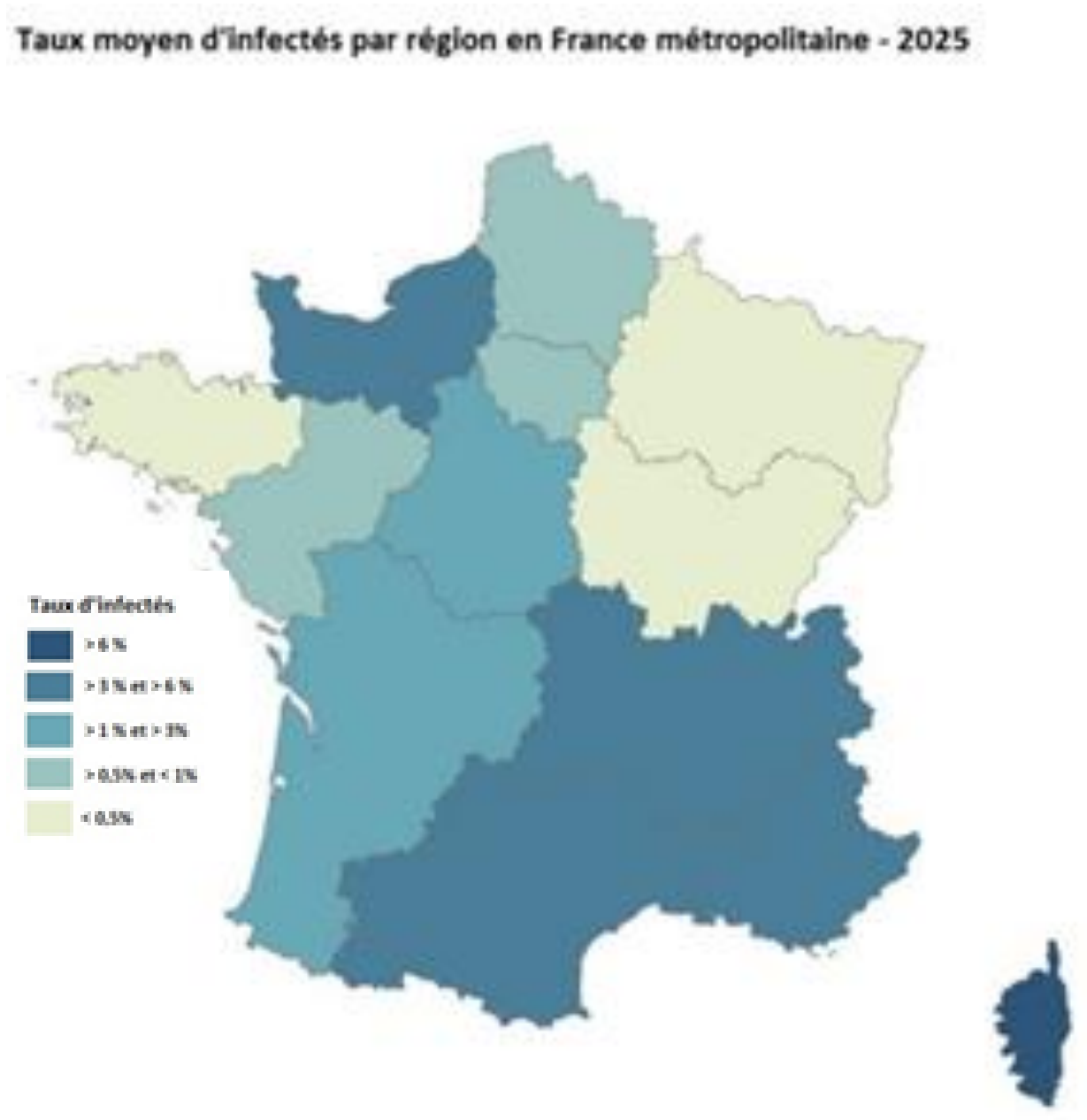

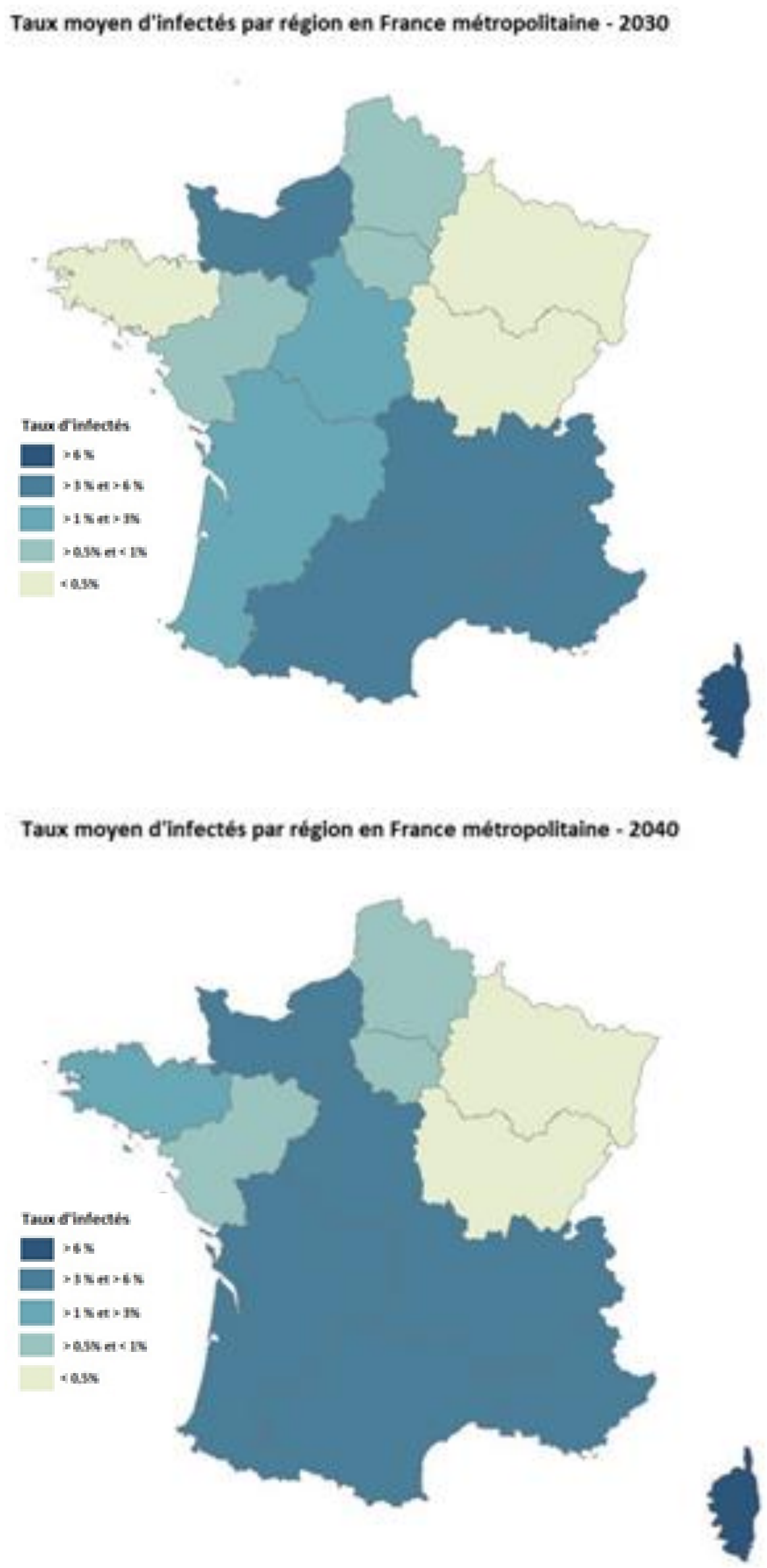

***Figure 3****. Probabilité de survenue d'épidémies dans chaque région selon le niveau de transmission assumée au cours du temps pour un scénario « fort ».*

Le deuxième indicateur considéré est le nombre de personnes ayant été infectées à la fin d'une saison. Nous avons considéré ici que tous les cas étaient recensés, ce qui est rarement le cas, en particulier pour les arboviroses qui sont souvent peu symptomatiques.

Ainsi, nous trouvons qu'en 2020, entre 100,000 et 1,270,000 personnes pourraient être infectées par an. Ce chiffre pourrait atteindre entre 250,000 et 2,000,000 de personnes en 2040 (figure 3).

### *2.2.1 Construction de scénarios pour les chocs de l'ACPR*

Sur l'horizon temporel de 2021 à 2050, la probabilité d'apparition d'épidémie a été estimée par région en intégrant les effets du réchauffement climatique et la colonisation du territoire de France Métropolitaine par le moustique tigre. Les estimations de personnes affectées par année et par région sont ensuite raffinées en précisant les principales conséquences d'une infection par la Dengue : le Décès dans 1 % des cas (WHO. 2017), un état de maladie menant à consulter dans 40 % des cas et une durée d'arrêt de travail d'environ 10 jours dans 1 % des cas. Ces pandémies sont traduites en sinistralité additionnelle (tableau 2 et 3).

| | | 2021-2024 | 2025-2029 | 2030-2039 | 2040-2049 | 2050 |
|---|---|---|---|---|---|---|
| **Granularité Nationale** | Facteur additif | 0,002 % | 0,002 % | 0,002 % | 0,002 % | 0,002% |
| | Facteur multiplicatif | 6,3% | 3,8% | 5,5% | 5,5% | |
| **Granularité par région** | | | | | | |
| **Auvergne-Rhône-Alpes** | Facteur additif | 0,005 % | 0,005 % | 0,005 % | 0,005 % | 0,005% |
| | Facteur multiplicatif | 0,00% | 0,3% | 0,3% | 0,3% | |
| **Bourgogne-Franche-Comté** | Facteur additif | 0,000 3% | 0,000 3% | 0,000 3% | 0,000 3% | 0,00 03% |
| | Facteur multiplicatif | 3,9% | 4,2% | 17,2% | 17,2% | |
| **Bretagne** | Facteur additif | 0,002 % | 0,002 % | 0,002 % | 0,002 % | 0,002% |
| | Facteur multiplicatif | 3,872 % | 3,033 % | 2,193 % | 2,193 % | |
| **Centre-Val de Loire** | Facteur additif | 0,003 % | 0,003 % | 0,003 % | 0,003 % | 0,003% |
| | Facteur multiplicatif | 0,6% | 11,0% | 2,2% | 2,2% | |
| **Corse** | Facteur additif | 0,02% | 0,02% | 0,02% | 0,02% | 0,02 % |
| | Facteur multiplicatif | 0,9% | 2,2% | 2,5% | 2,5% | |
| **Grand Est** | Facteur additif | 0,000 3% | 0,000 3% | 0,000 3% | 0,000 3% | 0,00 03% |
| | Facteur multiplicatif | 3,9% | 3,2% | 17,2% | 17,2% | |
| **Hauts-de-France** | Facteur additif | 0,001 % | 0,001 % | 0,001 % | 0,001 % | 0,001% |
| | Facteur multiplicatif | 0,0% | 1% | 3% | 3% | |
| **Ile-de-France** | Facteur additif | 0,001 % | 0,001 % | 0,001 % | 0,001 % | 0,001% |
| | Facteur multiplicatif | 0,0% | 1,3% | 2,5% | 2,5% | |
| **Normandie** | Facteur additif | 0,005 % | 0,005 % | 0,005 % | 0,005 % | 0,005% |
| | Facteur multiplicatif | 1,5% | 1,4% | 3,0% | 3,0% | |
| **Nouvelle-Aquitaine** | Facteur additif | 0,003 % | 0,003 % | 0,003 % | 0,003 % | 0,003% |
| | Facteur multiplicatif | 0,1% | 11,0% | 2,2% | 2,2% | |
| **Occitanie** | Facteur additif | 0,006% | 0,006% | 0,006% | 0,006% | 0,006% |
| | Facteur multiplicatif | 0,6% | 0,6% | 2,5% | 2,5% | |
| **Pays de la Loire** | Facteur additif | 0,001% | 0,001% | 0,001% | 0,001% | 0,001% |
| | Facteur multiplicatif | 0,0% | 4,2% | 2,5% | 2,5% | |
| **Provence-Alpes-Côte d'Azur** | Facteur additif | 0,005% | 0,005% | 0,005% | 0,005% | 0,005% |
| | Facteur multiplicatif | 1,4% | 1,3% | 6,8% | 6,8% | |

***Tableau 2****. Impact additif et multiplicatif sur le taux de mortalité par horizon de projection.*

Les différents éléments constituant le scénario sont :

Le Facteur additif correspond à une majoration additive des taux de mortalité annuels. *Eg. Un Facteur additif de 0,002 % des taux de mortalité fait qu'un taux de mortalité, avant choc de 0,03 % passe à 0,032% après choc*.

Le Facteur multiplicatif correspond à une aggravation annuelle du décalage des tables de mortalité. *Eg. Une Facteur multiplicatif de 2 % fait que les taux de mortalité qui décalent de 0,002 % la première année, décalent de 0,002 % x 1,02 la seconde année, de 0,002 % x 1,02 x 1,02 la seconde année …*

| | | 2021-2024 | 2025-2029 | 2030-2039 | 2040-2049 | 2050 |
|---|---|---|---|---|---|---|
| **Granularité Nationale** | Consultation / Urgence | 0,7911% | 1,0407% | 1,2408% | 1,5808% | 1,9208% |
| | ITT J | 0,0198% | 0,0260% | 0,0310% | 0,0395% | 0,0480% |
| **Granularité par région** | | | | | | |
| **Auvergne-Rhône-Alpes** | Consultation / Urgence | 2,1094% | 2,1094% | 2,1441% | 2,1788% | 2,2135% |
| | ITT J | 0,0527% | 0,0527% | 0,0536% | 0,0545% | 0,0553% |
| **Bourgogne-Franche-Comté** | Consultation / Urgence | 0,1221% | 0,1458% | 0,1694% | 0,3154% | 0,4613% |
| | ITT J | 0,0031% | 0,0036% | 0,0042% | 0,0079% | 0,0115% |
| **Bretagne** | Consultation / Urgence | 0,0611% | 0,5985% | 0,9272% | 1,0288% | 1,1305% |
| | ITT J | 0,0015% | 0,0150% | 0,0232% | 0,0257% | 0,0283% |
| **Centre-Val de Loire** | Consultation / Urgence | 1,1600% | 1,1970% | 1,8543% | 2,0577% | 2,2611% |
| | ITT J | 0,0290% | 0,0299% | 0,0464% | 0,0514% | 0,0565% |
| **Corse** | Consultation / Urgence | 8,6167% | 9,0000% | 10,0000% | 11,2629% | 12,5259% |
| | ITT J | 0,2154% | 0,2250% | 0,2500% | 0,2816% | 0,3131% |
| **Grand Est** | Consultation / Urgence | 0,1221% | 0,1458% | 0,1694% | 0,3154% | 0,4613% |
| | ITT J | 0,0031% | 0,0036% | 0,0042% | 0,0079% | 0,0115% |
| **Hauts-de-France** | Consultation / Urgence | 0,3221% | 0,0320% | 0,3904% | 0,4400% | 0,4896% |
| | ITT J | 0,0081% | 0,0008% | 0,0098% | 0,0110% | 0,0122% |
| **Ile-de-France** | Consultation / Urgence | 0,3221% | 0,0134% | 0,3904% | 0,4400% | 0,4896% |
| | ITT J | 0,0081% | 0,0003% | 0,0098% | 0,0110% | 0,0122% |
| **Normandie** | Consultation / Urgence | 2,0800% | 2,2400% | 2,4000% | 2,7600% | 3,1200% |
| | ITT J | 0,0520% | 0,0560% | 0,0600% | 0,0690% | 0,0780% |
| **Nouvelle-Aquitaine** | Consultation / Urgence | 1,1899% | 1,1970% | 1,8543% | 2,0577% | 2,2611% |
| | ITT J | 0,0297% | 0,0299% | 0,0464% | 0,0514% | 0,0565% |
| **Occitanie** | Consultation / Urgence | 2,2780% | 2,3484% | 2,4188% | 2,7250% | 3,0313% |
| | ITT J | 0,0570% | 0,0587% | 0,0605% | 0,0681% | 0,0758% |
| **Pays de la Loire** | Consultation / Urgence | 0,3221% | 0,1458% | 0,3904% | 0,4400% | 0,4896% |
| | ITT J | 0,0081% | 0,0036% | 0,0098% | 0,0110% | 0,0122% |
| **Provence-Alpes-Côte d'Azur** | Consultation / Urgence | 2,0276% | 2,1733% | 2,3190% | 3,1108% | 3,9027% |
| | ITT J | 0,0507% | 0,0543% | 0,0580% | 0,0778% | 0,0976% |

***Tableau 3****. Impact additif et multiplicatif sur les frais de soin et les Arrêts de Travail par horizon de projection.*

## Remerciements

Nous remercions tout particluièrement Gérard DUVALLET, Entomologiste médical et Professeur émérite à l'université de Montpellier pour son aide lors de l'élaboration de ce projet.

## Bibliographie